\documentclass[12pt]{article}
\input psfig.tex
\usepackage{graphicx}
\usepackage{amssymb}
\usepackage{epstopdf}
\newcommand{\ignore}[1]{}
\newcommand{\be}{\begin{equation}} \newcommand{\ee}{\end{equation}}
\newcommand{\ba}{\begin{eqnarray}} \newcommand{\ea}{\end{eqnarray}}
\newcommand{\nn}{\nonumber} \renewcommand{\bf}{\textbf}
\newcommand{\ra}{\rightarrow}

\renewcommand{\a}{\alpha} \renewcommand{\b}{\beta}

\newcommand{\p}{\partial}  
  \newcommand{\NN}{\vec \nabla}  

\begin{document}

\input{epsf}

\title{ The Dynamical Mixing of Light and Pseudoscalar Fields} 

\author{Sudeep Das$^a$, Pankaj Jain$^b$, John P. Ralston$^c$ and Rajib
Saha$^b$\\
$^a$Department of Astrophysical Sciences\\ Princeton University \\
New Jersey 08544, USA\\
$^b$Physics Department, IIT, Kanpur - 208016, India\\
$^c$Department of Physics \& Astronomy \\
University of Kansas\\ Lawrence, KS-66045, USA\\
}
\maketitle

\abstract{\small We solve the general problem of mixing of electromagnetic and scalar or pseudoscalar fields coupled by axion-type interactions ${\cal L}_{int} =g_{\phi}  \phi \, \epsilon_{\mu \nu \a \b}F^{\mu 
\nu}F^{\a \b} $. The problem depends on several dimensionful scales, including the magnitude and direction of background magnetic field, the pseudoscalar mass, plasma frequency, propagation frequency, wave number, and finally the pseudoscalar coupling.  We apply the results to the first consistent calculations of the mixing of light propagating in a background magnetic field of varying direction, which shows a great variety of fascinating resonant and polarization effects.  }

\medskip
\medskip

For about 20 years the mixing of light and pseudoscalar fields in propagation 
has been studied with fascination 
\cite{Karl}-\cite{CG94}.  
The subject generated renewed attention 
in the context of cosmological observables that can probe exceedingly 
small couplings \cite{raffelt,Brockway,Rosenberg,PJ}.   
One recent approach proposes that the dimming of 
supernova light might be explained by transition of light into unobserved
pseudoscalar, or ``axion,'' modes \cite{Csaki}, although this effect might
be limited by observations of radio galaxies \cite{Bassett}.  
It has also been pointed out that pseudoscalar field can generate magnetic
fields due to their coupling with photons \cite{Ng}.
Polarization observables are even more sensitive than intensity: 
for coupling constants many
orders of magnitude too 
small to cause dimming, the cumulative evolution of phase shifts can generate phenomena clearly violating the Maxwell 
equations in plasmas\cite{ourselves}.  Several laboratory experiments 
have also sought the spontaneous resonant conversion of dark matter 
axions to photons, and explored the possibilities of conversion in 
lab-made magnetic fields.  

There is a well-established theoretical technology of mixing light with 
a background magnetic field transverse to propagation.  Yet despite long 
study, we know of no complete solution to the mixing problem depending 
on every possible variable.  And there is no wonder, as there are many 
dimensionful scales, including the magnitude and direction of background 
magnetic field, the pseudoscalar mass, plasma frequency, propagation 
frequency, wave number, and finally the pseudoscalar coupling.  By 
approaching the problem with new methods here, we will be able to survey 
various limits used in the literature and also present a convincing 
resolution of the dynamics in a slowly varying background field of arbitrary 
direction. 

The basic Lagrangian assumes a pseudoscalar\footnote{We may let $\phi$ also be a scalar field given parity violation. } field $\phi$ coupled to the electromagnetic 
field strength $F_{\mu \nu}$ by the action \ba S  = \int d^{4}x 
\sqrt{-g}  \bigg[ - \frac{1}{4} F_{\mu \nu}F^{\mu \nu} + \sqrt{g} \, g_{\phi}  \phi \, \epsilon_{\mu \nu \a \b}F^{\mu 
\nu}F^{\a \b} \label{Ldefined}  \\ +j_{\mu}A^{\mu} + \frac{1}{2}\p_{\mu} 
\phi \p_{\mu}\phi -{1 \over 2}m_{\phi}^{2}\phi^{2} 
-V(\phi)\bigg]. \label{Ldefined2}  \ea 
We include a coupling to a current $j_{\mu}$ for completeness.  For the 
purposes of linear propagation the potential $V(\phi)$ can be ignored 
as a small perturbation, and the metric $g$ replaced by a given background 
form.  Certain non-local plasma effects, described by the plasma frequency, 
Faraday rotation, etc., may also need to be incorporated. 

By translational symmetry, certain eigenmodes will evolve like $e^{i k_{i}z}$ 
in propagation over a distance $z$, where $k_{i}$ are wave numbers to be 
determined.  This is simple and obvious.  Yet one might claim the opposite 
that $k$ should be fixed, while frequency $\omega$ remains to be determined, 
as so common in quantum mechanics and neutrino oscillations.  Indeed some 
literature solves for $\omega$ eigenvalues without discussion.  Physics 
is local, and by Huygen's principle, i.e. the use of causal Green functions,  
a source with known time dependence $e^{ - i \omega t}$ must over distance 
develop its own wave numbers $k_{i}$ to propagate on-shell.  There are no 
boundary conditions of fixed $k$ from which to calculate frequency, so it 
is not negotiable that $k_{i}$ are to be solved, just as in careful work 
on neutrino oscillations \cite{lipkin,stodolsky}.   
 We also give extra attention to 
maintaining gauge invariance, which we have not seen before. 
The physics turns out to be surprisingly 
intricate.

We apply the revised propagation equations to the interesting problem 
of light traveling in a background magnetic field of varying direction. 
  For the parameter values of axion masses, magnetic fields and 
  couplings commonly assumed, the magnitude of new changes is often 
non-negligible.  This fact aside, the results themselves are fascinating, 
  and full of remarkable complexity and structure, somewhat like a 
  generalized version of the resonant propagation of neutrinos.   We 
  think this is very interesting: The possible existence of axions can be 
  probed in polarization observables for parameters ranges far smaller 
  than will cause a dimming of light by direct conversion.  Although 
  axion-related dimming is given some credence it is usually assumed 
  there are no exotic polarization effects to be observed.  We find that 
  the {\it absence}  of exotic polarization effects would be able to rule 
  out the light-dimming hypothesis.  Confrontation with data on 
  polarization, of course, needs a detailed study of many potential 
  backgrounds to any signal, and would go beyond the scope of this paper. 
     Our main task is simply to get the propagation equations resolved 
     once and for all.

\section{Gauge Invariant Methods}

\subsection{Equations for $E$ and $\phi$} 
To eliminate difficulties of gauge invariance we first obtain the non-covariant form of the Maxwell 
equations with no approximations \cite{mohanty}:  \begin{eqnarray}
\nabla\cdot {\vec E} &=& g_{\phi}\nabla\phi\cdot{(\vec {\cal B}+ \vec B )} +\rho ;\label{emeq1} 
\\
\nabla\times{\vec E} + {\partial {(\vec {\cal B}+ \vec B )}\over \partial t} &=& 0 ;\label{emeq2} \\
 \nabla\times{\vec B} - {\partial {\vec E}\over \partial
t}
&=& g_{\phi} \left({\vec E}\times \nabla \phi - {  (\vec {\cal B}+ \vec B )} 
{\partial \phi\over \partial t} 
\right)  + \vec j ; \label{emeq3} 
\\
\nabla\cdot{  (\vec {\cal B}+ \vec B )}&=& 0.
\label{emeq4}
\end{eqnarray}  Here ${\cal B}_{i}+B_{i} =\frac{1}{2}\epsilon_{ijk}F^{jk}$ and 
$E_{i}=F^{0i}$ are the usual magnetic and electric fields. 
Here $\vec{\cal B}$ and $\vec B$ represent the magnetic field due to
the background and due to the electromagnetic wave respectively. 

In anticipation we note that the revised ``Gauss's Law'' Eq. \ref{emeq1} couples the longitudinal electric field to $\NN \phi$.  This creates a qualitative change compared to light in free space, where the longitudinal mode does not normally propagate.  If $\phi$ propagates we now have a propagating longitudinal light field.  If there is a plasma, then the ordinary Gauss's Law becomes $\NN \cdot \epsilon E =\rho_{free}$, where $\epsilon$ is the dielectric constant (or ``permitivity``).  Since $\epsilon= \epsilon(\omega)$ is not local in the time domain, we will incorporate it below in the Fourier-transformed equations.

The pseudoscalar field's equation of motion is
\begin{equation}
{\partial^2 \phi\over \partial t^2} - \nabla^2\phi + m_\phi^2\phi = 
-g_{\phi}{\vec
E}\cdot
(\vec {\cal B}+ \vec B )
\label{pseudoscalar}
\end{equation}  Gauge invariance is explicit, and one can check current conservation directly, $$ \NN \cdot \vec j+ {\p \rho \over \p t}=0. $$ 

 Assume $\vec {\cal B}$ solves the zeroeth order Maxwell equations with no $\phi$ background.  The linearized equations for $\vec E/c<< \vec {\cal B}, \, \vec B <<\vec {\cal B}$ are  \begin{eqnarray}
\nabla\cdot {\vec E} &=& g_{\phi}\nabla\phi\cdot{\vec {\cal B} } +\rho; \label{remeq1} 
\\
\nabla\times{\vec E} + {\partial \vec  B \over \partial t} &=& 0; \label{remeq2} \\
 \nabla\times{\vec B} - {\partial {\vec E}\over \partial
t}
&=& -g_{\phi}     \vec {\cal B}   
{\partial \phi\over \partial t} 
   + \vec j ; \label{remeq3} 
\\
\nabla\cdot{ \vec B }&=& 0. \label{remeq4}
\end{eqnarray}  

Proceed to get a wave equation for $\vec E$ by taking the curl of Faraday's 
Law, $$  \NN \times \NN \times \vec E = -\NN^{2} \vec E + \NN \NN \cdot 
\vec E = -{\p \over \p t } \NN \times  \vec B, $$  and substituting Eqs. 
\ref{emeq1}, \ref{emeq3}.  Replacing $\vec B \sim \vec {\cal B}$ gives  
\ba  
-\NN^{2 }\vec E + {\p^{2} {\vec E} \over \p t^{2}  } = g_{\phi} \vec 
{\cal B}{\p^{2} \phi \over \p t^{2}  }-g_{\phi}  \NN ( \NN\phi \cdot \vec 
{\cal B} ).  \label{Ewave} \ea  In this equation the longitudinal part 
of $\vec E$ mixes with $ \NN\phi $.  Take the transverse(sub-$T$) and 
longitudinal parts (sub-$L$) of the electric wave equation, for wave 
number $\vec k$, with $E_{L}= \hat k \cdot \vec E$: \ba  (\,   k^{2} 
-\omega^{2} \,)E_{T}=  -g_{\phi}\omega^{2}  {\cal B}_{T} \phi ; 
\label{Ewave2} \\  (\,   k^{2} -\omega^{2} \,)E_{L}=  g_{\phi} 
(k^{2}-\omega^2)   {\cal B}_{L} \phi .   \ea  
There clearly exists no gauge in which the longitudinal electric field 
decouples from the problem.  
If we limit the study to $\NN\phi \cdot \vec {\cal B} =0$, then Gauss' Law 
makes $\vec E$ transverse.  Everything in the literature is perfectly 
consistent.  

\subsection{Equations for $D$ and $\phi$}

Another method is needed 
when $\vec k \cdot \vec {\cal B}  \neq 0$.
Many linearized electromagnetic theories can be encompassed by the 
equations:
  \begin{eqnarray}
\NN \cdot {\vec D} &=& 0; \label{deq1} \\
\nabla\times{\vec  E} + { \partial  \vec B \over \partial t} &=& 0  ;
\label{deq2} \\
  \nabla\times{\vec H} - {\partial \vec  D \over \partial t}
&=& 0 ; \label{deq3} \\
\NN \cdot{\vec B}&=& 0.  \label{deq4}
\end{eqnarray}  The purpose of the ``archaic'' representation via 
$\vec D$ is to have a field which is perfectly transverse.   With $\vec D$ 
the transverse wave operator is greatly simplified: $$ \NN \times 
(\NN \times  \vec D ) \ra  -\NN^{2} \vec D.  $$  This effectively reduces 
the freedoms of the propagating gauge fields from 3 to 2: one would have 
to use 4-state mixing of 3 $\vec E$ components and one $\phi$ if this were 
not arranged. 

Can we make $\vec D$ and $\vec H$ serve in Eqs. \ref{remeq1} -\ref{remeq4}, and also include plasma effects?  
We find Eqs. \ref{deq1}-\ref{deq4} consistent with the definitions:  \ba 
\vec D &=&  \epsilon \vec E - g_{\phi}  \phi \vec{\cal B}; \\ 
   \vec H  &=& \vec B\ .  \ea  The asymmetry here comes from having a magnetic background.
In our work we will assume the contribution to $\epsilon$ due to the plasma frequency
$\omega_p$, via \ba \ \ \ \epsilon = \left(1-{\omega_p^2\over \omega^2}\right) .\nn \ea 

\subsection{Decoupling}

From Faraday's Law and the $\vec D$ equation we have 
\ba {1\over 1-(\omega_p^2/ \omega^2)}\NN \times 
(\vec D+ g_{\phi} \vec {\cal B} 
\phi) =-{\p \vec B  \over \p t }=-{\p \vec H  \over \p t }; \nn \\  
\NN \times \NN \times (\, \vec D+ g_{\phi} \vec {\cal B} \phi \, ) = 
\left(1-{\omega_p^2\over \omega^2}\right)  {\p^{2}\vec D \over \p t^{2} }. 
\label{Deqn} \ea  
Together with the $\phi$ propagation from Eq. \ref{pseudoscalar},  the equations have been simplified as much as generally possible: the coupled system of $\phi, \, D_{x}, \, D_{y}, \,D_{z} $ have one locally decoupled mode, no longitudinal mode, and are equivalent to two coupled pde's with no approximations other than linearization.

We now drop terms of order $\NN {\cal B}/B$ as negligible compared to 
other length scales, including the splitting of modes, setting up the 
usual adiabatic limit.   We seek local plane wave solutions with 
$\NN \ra i \vec k$.  
The component of $\vec D$ perpendicular to $ \vec {\cal B}$ decouples: 
\ba (\, k^{2 } + \omega_{p}^{2} -\omega^{2}  \,)  \vec D \times \hat {\cal B}
=0.  \ea  
The other transverse projection of the $\vec D$ wave equation becomes 
\ba  (\,  k^{2 } + \omega_{p}^{2} -\omega^{2} \, ) \vec D \cdot 
 \, \hat {\cal B}_{T}  + k^{2 } g_{\phi}{\cal B}_{T} 
\phi =0.  \label{k2phi}\ea 
Notice that in using $\vec D$ the equation of motion, involving the curl 
of $\vec B$, is not used:  in fact it is satisfied as an identity.  
Conversely, when Faraday's Law is substituted into the $\vec E$ wave 
equation, then Faraday's Law is satisfied as an identity, and the equation 
of motion is solved (Eq. \ref{Ewave} ).   By subtracting Eq. \ref {k2phi} 
from the (in principle) independent wave Eq. \ref{Ewave} for $\vec E$ 
at the compatible point, we obtain a nice consistency check.
   
We turn to the coupled system: \ba   (\,  k^{2 } + \omega_{p}^{2} -\omega^{2} 
\, )\, \vec D \cdot \, \hat {\cal B}_{T}  + k^{2 } \, 
g_{\phi}{\cal B}_{T} \phi =0;  \\ 
{g_{\phi}{\cal B}_{T}\over 1-\omega_p^2/\omega^2 }\,\vec D \cdot  
\hat {\cal B}_{T}   + \left(\, k^{2}+m_{\phi}^{2} - \omega^{2} + 
{g_{\phi}^2 {\cal B}^{2}\over  1-\omega_p^2/\omega^2 }  \,\right) \,  
\phi =0. \ea 
The system can be solved directly for the dispersion relation 
$k^{2}=k^{2}(\omega)$ by setting to zero the determinant of the corresponding 
matrix $M$, defined by
$$ M\left(\matrix{\vec D \cdot \, \hat {\cal B}_{T}\cr \phi\cr }\right)=0\ . $$
However the eigenvalues $k^{2}$ needed are not on the diagonal.  Moreover $M$ is not symmetric, and non-symmetric matrices have eigenvectors which are not orthogonal.  

Much the same occurs in optics \cite{BornWolf}, where the corresponding equations for propagation with a tensor dielectric constant $\epsilon_{ij}$ are: \ba (\,  
k^{2}\delta^{T}(k)\epsilon^{-1} -  \omega^{2} \,) D=0; \\ \delta_{ij}^{T}(k)= \delta_{ij} -\hat k_{i}\hat k_{j}. \nn \ea  One seldom finds $\delta^{T}(k)\epsilon^{-1}$ to be symmetric.  Yet since $$\delta^{T}(k)D=D,$$ multiplication on the left by $\delta^{T}(k)$ yields a symmetric eigenvalue equation:
\ba (\,  k^{2}\delta^{T}(k)\epsilon^{-1}\delta^{T}(k) - 
\omega^{2} \,) D=0.\ea  The propagation eigenstates are obtained from the $2 \times 2$ matrix $\delta^{T}(k)\epsilon^{-1}\delta^{T}(k)$ in the sector transverse to $\vec k$.  This is considerably more subtle than (say) diagonalizing $\epsilon_{ij}$ first, and simply taking a transverse part.   

This indicates that further transformations are needed for a useful solution. 

\subsection{Orthogonal Modes}

First, $\hat D \times \hat {\cal B}$ 
decouples from $\phi$ and propagates like ordinary light (including 
plasma frequency) with wave number $k_{0} =\sqrt{ \omega^{2}-\omega_p^2}$.  

We made the rest of the transformation by inspection.  Define 
\ba \bar \phi = k_0  \phi; \nn \\ 
 \bar D  = {D\cdot \hat {\cal B} +  g {\cal B}_T  \phi \over 
 \sqrt{1-(\omega_p^2/\omega^2)}}\   . \label{clever} \ea  
 Now the propagation matrix is symmetric and eigenvalue $k^{2}$ 
 lies on the diagonal:  \ba  \left(\begin{array}{cc}  k^{2}+ \omega_{p}^{2} 
 -\omega^{2} &  g_{\phi}{\cal  B}_T \omega   \\  g_{\phi}{\cal  
 B}_T \omega & k^{2}+ \tilde m_{\phi}^{2} -\omega^{2}  
 \end{array}\right) \left(\begin{array}{c} \bar  D  \\ \bar \phi  
 \end{array}\right)  \ea with 
 \ba \tilde m_{\phi}^2  = m_{\phi}^2  + {g_{\phi}^2 {\cal B}_L^2
 \over 1-(\omega_p^2/\omega^2)}\ . \ea  
As a consequence propagation generates 
 unitary rotations of $ ( \bar D , \, \bar \phi )$.  Go to a new basis 
 \ba |\eta> =O |\psi_{\Lambda}> ; \:\:\:\:  O = \left(\matrix 
{\cos\bar \theta & -\sin\bar \theta\cr \sin\bar \theta & \cos \bar \theta } 
\right)\ .\ea  
The mixing angle diagonalizing propagation is \ba  \tan 2 \bar \theta =  \frac{g_{\phi } \omega {\cal B}_{T}  } {\tilde m_{\phi}^{2} -\omega_{p}^{2}}. 
\label{goodtan} \ea 
    The dispersion relations are \ba 
 k_{1}^{2} &=&  \omega^2 -{1\over 2} (\, \tilde  m_{\phi}^{2} + \omega_{p}^{2}) 
 - {1\over 2} \sqrt{ \Omega^{4}} ;  \\
k_{2}^{2} &=& \omega^2  -{1\over 2} (\tilde m_{\phi}^{2} + \omega_{p}^{2} ) +   {1\over 2} \sqrt{ \Omega^{4}}, \ea where \ba 
 \Omega^{4}= 4 g_{\phi}^{2}{\cal B}_{T}^{2} \omega^{2} +(\tilde m_{\phi}^{2} -\omega_{p}^{2})^{2}    .  \ea  
 
By inspection of these results, the eigenvalues and mixing are just the 
same as solving the ${\cal B}_{L}=0$ limit and making the replacement 
$m_{\phi}^{2} \ra  \tilde m_{\phi}^{2}= m_{\phi}^{2}+ g_{\phi}^{2} 
{\cal B}_L ^{2}/[1-(\omega_p^2/\omega^2)]. $   

\subsubsection{Plane Wave Simplification} 

There are circumstances where 
neglecting $\NN {\cal B}/{\cal B}$ may be not possible. Then Eq. \ref{Deqn} and Eq. \ref{pseudoscalar} cannot be simplified further.  However if 
the propagation can be reduced to plane wave modes with constant parameters, 
there is a simple way to understand the modes. 

First solve the longitudinal mode using Gauss' Law: 
\ba \vec k \cdot { \epsilon \vec E} &=& g_{\phi} (\, \vec k\phi \,) \cdot \vec 
{\cal B} ; \nn \\  \epsilon \vec E_{L} &=& \hat k  \, g_{\phi} \hat k\phi\cdot 
\vec {\cal B}.  \label{EL} \ea 
Here $\hat k = \vec k/k$ is a non-local operator.  Insert the solution where it appears in the propagation of $\phi$, 
Eqn. \ref{pseudoscalar}: 
\ba ( \, \omega^{2} +k^{2 }+ m_{\phi}^{2} \,) \phi & =& -g_{\phi}{\vec
E}\cdot \vec {\cal B} , \nn \\ & \ra &  -g_{\phi}E_{T} {\cal B}_{T} -
{ g^{2}{\cal B}_L^{2}\phi \over \epsilon} 
\label{pseudoscalar2} \ea 
 Observe that the effects on equations for $\phi$ are the same as replacing 
$m_{\phi}^{2} \ra  \tilde m_{\phi}^{2}=m_{\phi}^{2}+ g_{\phi}^{2}{\cal B}_L^{2}/\epsilon
. $   Meanwhile the {\it transverse} projection of the electric equation, 
Eq. \ref{Ewave}, also involves only $E_{T}$ and $\phi$.  Since this subsystem 
has decoupled, they must have modes which are linear combinations of $\phi$ 
and $E_{T}$:  finally we recover the transformation to reveal that 
$\bar D = E_{T}$ in this limit. 

\subsubsection{Non-Perturbative Effects, and A New Resonance}
 
The axion mass must be very small, and it often appears safe to take 
$m_{\phi} \ra 0, \,  \omega_{p} \ra 0$ smoothly. This is the assumption 
of several phenomenological applications.  Let us revisit that question 
in light of the mixing solution Eq. \ref{goodtan}, which reduces to 
\ba  \tan 2 \bar \theta  =  {{ \cal  B}_T  \omega  \over g_{\phi} 
{\cal  B_{L} }^{2}}, \:\:\:   m_{\phi} = \omega_{p}= 0.  \label{inverse} \ea  
This is a very interesting result: the mixing is {\it inversely} proportional 
to the coupling constant, a typical non-perturbative effect.  

Can we take Eq. \ref{inverse} seriously?   In current cosmological 
propagation applications, the order of scales is $\omega_{p}^{2} > 
m_{\phi}^{2} > g_{\phi} ^{2}{\cal  B }^{2}$, and Eq. \ref{inverse} does 
not apply.  However in lab experiments we can control the vacuum to 
make $\omega_{p} \sim 0$ to a very high degree.  

\begin{figure}
 \centering

  \epsfxsize=4.5in \epsfysize=2.5in
\epsfbox{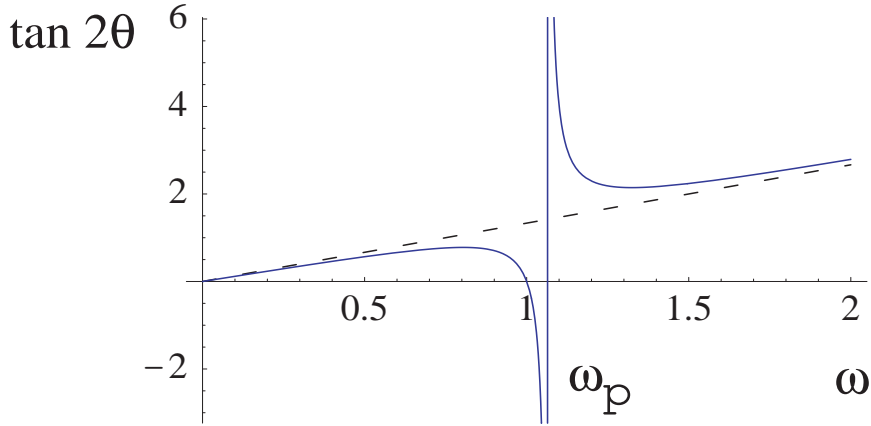}

\caption{ Behavior of $tan(2 \bar \theta)$ in the vicinity of $\omega \sim \omega_{p}$ for $ m_{\phi}^{2} \sim \omega_{p}^{2}$.  The dashed line shows the calculation for ${\cal B_{L}}=0$.  Parameter values have been rescaled to make the resonance visible: it would be exceedingly narrow given current beliefs for $g_{\phi }$.    }

\label{fig:ResonantTangent}

\end{figure}

Restoring the full dependence on all variables, $tan(2\bar \theta)$ of Eq. \ref {goodtan} is compared to the traditional formula valid for ${\cal B_{L}}=0$ in Fig. \ref{fig:ResonantTangent}.  There is a type of resonance at $\omega = \omega_{p}$, with width $$\Delta\omega \sim \omega_{p} {g_{\phi}^{2}{\cal B_{L}}^{2}  \over 2(m_{\phi}^{2} -\omega_{p}^{2} )}. $$  Given current values of $g_{\phi}$, the resonance may be too narrow to observe.  

The resonant effect is very interesting conceptually.  Qualitatively, at 
resonance it appears that the longitudinal mode of a plasma oscillation 
becomes very strongly mixed with the pseudoscalar field, depending on the 
difference of masses. As we mentioned earlier, $\phi$ mixes with $\vec 
E_{L} $: indeed due to the constraint of Gauss's Law, it is the {\it same 
dynamical phenomenon} as the longitudinal field.   Let us estimate some 
magnitudes: when fully mixed, $\bar \phi \sim \bar D  $, or $$ \phi \sim 
{E_{T} \over \omega_{p}} .$$  As mentioned earlier $E_{L}\sim { g_{\phi} 
\phi {\cal B_{L}} \over \epsilon} $. Together the relations predict \ba 
{E_{L} \over E_{T}} \sim  { g_{\phi}{\cal B_{L}} \over \epsilon \omega_{p}} 
\sim { g_{\phi}{\cal B_{L}} \omega_{p} \over \omega_{p}^{2}- \omega^{2}} .\nn 
\ea Thus there is always a frequency for which we may observe the formerly 
non-interacting pseudoscalar electromagnetically, and as a form of {\it 
longitudinally polarized light }: The $E_{L}$ being observable and affecting 
instruments just as much as a longitudinal field in a plasma oscillation.  
Given sufficiently fine measurements the ``invisible axion'' could in 
principle be ``visible.''

We hope to explore more deeply the potential laboratory repercussions of these phenomenon in another paper.  Given that most current interest centers on cosmological propagation, we turn to studying the effects of a varying $\vec{\cal B}$ field in the next Section. 

\section{Three Mode Mixing: Varying $\vec{\cal B}$}

We next consider the adiabatic propagation of light through a
background magnetic field which varies slowly in direction. This problem has
not been solved before. The results are far from trivial, and give substance 
to many cosmological applications assuming some ``fluctuating'' magnetic fields with typical coherence lengths.  As we will show, the variety of physical phenomena one can observe is very great. In some limits,  writing a transition probability and taking a statistical average may suffice, but in other limits the polarization effects are quite spectacular.  The dynamical possibilities 
for the mixing of light actually exceed those for neutrino-mass mixing, 
which has been studied for nearly 50 years and still appear 
inexhaustible. Indeed we were engaged in the general propagation problem 
by this very attractive method to probe the existence of light pseudoscalars 
in cosmology \cite{ourselves}.

The physically observable density matrix $\rho$ is given by  \ba & \rho&=   
\left(\begin{array}{cc} <\,E_{||} E_{||}^{*} \,>   & <\, E_{||}E_{\perp}^{*} \,> \\<\, 
E_{\perp} E_{||}^{*} \,> &<\, E_{\perp} E_{\perp}^{*}  \,>  \end{array}\right) , \ea where $<\: >$ denotes the statistical averages occurring in propagation\footnote{We decline to develop a $3 \times 3$ density matrix including the longitudinal mode, as unlikely to be observed in these circumstances}.  

Orient the z-axis along the direction of the wave.  Let angle $\xi$ measure the
direction of the background field relative to the x-axis: 
\begin{equation}
 \vec {\cal B_T} = B\cos\xi(z) \hat i + B \sin\xi(z) \hat j
\end{equation}
We fix the magnitude of the background 
magnetic field to identify effects arising due to varying magnetic field 
direction. A changing background magnetic field magnitude is easily included
in the formalism. For the same reason we ignore the variation in plasma density
along the path.  The effects of varying plasma density 
for fixed field direction has been studied in detail elsewhere \cite{ourselves}.

The wave equation can now be written as
\begin{equation}
(\omega^2 + \partial_z^2)\left(\matrix{A_x\cr A_y\cr \phi\cr}\right)
- \left(\matrix{\omega_p^2 & 0 & -gB\omega \cos\xi\cr
                0  & \omega_p^2 & -gB\omega \sin\xi\cr
		 -gB\omega \cos\xi & -gB\omega \sin\xi & m_\phi^2\cr}\right)
		 \left(\matrix{A_x\cr A_y\cr \phi\cr}\right)
		 =0
\label{eq:wave_varyB}
\end{equation}
where $\vec A=\vec E/\omega$. We dropped $g^{2}{\cal B}_L^{2}$ terms as negligible
for intergalactic propagation with typical parameters. With a slowly varying background and working in the adiabatic limit, we define transformed fields $A'_x$, $A'_y$ and $\phi'$ such that
\begin{equation}
\left(\matrix{A_x\cr A_y\cr \phi\cr}\right) = 
\left(\matrix{\cos\beta & -\sin\beta & 0 \cr
                \sin\beta  & \cos\beta & 0   \cr
		0 & 0 & 1\cr}\right)
		\left(\matrix{A'_x\cr A'_y\cr \phi'\cr}\right)
\label{eq:three_mixing}
\end{equation}
The wave equation reduces to 
\begin{equation}
(\omega^2 + \partial_z^2)\left(\matrix{A'_x\cr A'_y\cr \phi'\cr}\right)
- \left(\matrix{\omega_p^2 & 0 & 0\cr
                0  & \omega_p^2 & -gB\omega\cr
		 0 & -gB\omega  & m_\phi^2\cr}\right)
		 \left(\matrix{A'_x\cr A'_y\cr \phi'\cr}\right)
		 =0.
\end{equation}
Here $\beta = \xi - \pi/2$. The equation reduces to the 
case of two component mixing which can be solved along the lines discussed
in Ref. \cite{ourselves}. Once we have obtained all the correlators between  
$A'_x$ and $A'_y$, we can express the required correlators as
\begin{eqnarray}
<A_x^*(z) A_x(z)> &=& \sin^2\xi(z) <A'^*_x(z)A'_x(z)> + \cos^2\xi(z)
<A'^*_y(z)A'_y(z)>\nonumber\\ 
&+& \cos\xi(z) \sin\xi(z)\ \left(<A'^*_x(z)A'_y(z)>
+ <A'^*_y(z)A'_x(z)>\right)\nonumber\\
<A_y^*(z) A_y(z)> &=& \cos^2\xi(z) <A'^*_x(z)A'_x(z)> + \sin^2\xi(z)
<A'^*_y(z)A'_y(z)>\nonumber\\ 
&-& \cos\xi(z) \sin\xi(z)\ \left(<A'^*_x(z)A'_y(z)>
+ <A'^*_y(z)A'_x(z)>\right)\nonumber\\
<A_x^*(z) A_y(z)> &=& -\cos\xi(z)\sin\xi(z) \left(<A'^*_x(z)A'_x(z)> 
-  <A'^*_y(z)A'_y(z)>\right)\nonumber\\ 
&+& \sin^2\xi(z) <A'^*_x(z)A'_y(z)>\nonumber\\
&-&\cos^2\xi(z) <A'^*_y(z)A'_x(z)>\ .
\end{eqnarray}
The correlators appearing on the right hand side of these equations
can be calculated by using the results in Ref. \cite{ourselves}.

\begin{figure}
\hskip -0.8in
\psfig{file=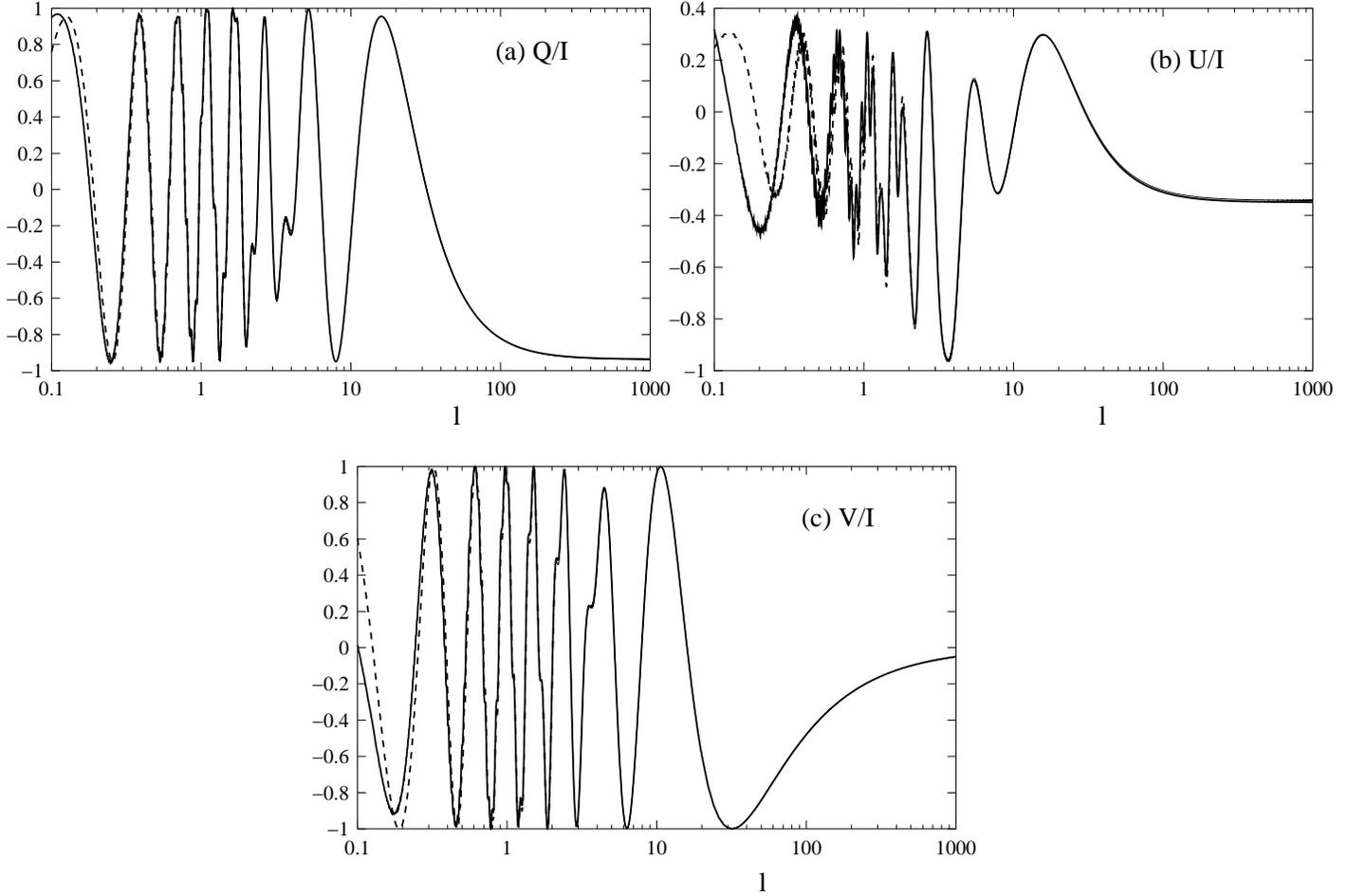}
\caption{ 
Normalized Stokes parameters (a) $Q/I$, (b) $U/I$ and (c) $V/I$ as a 
function of the length parameter $l$
for varying direction of background magnetic field;
 the magnitude
 $|\vec {\cal B}|$ and $\omega_{p}$ are constant.
Curves generated by direct numerical integration (solid)  and adiabatic 
analytic calculation(dashed). 
Parameters $gB=1.0$, $L=100$,
$m_\phi^2/\omega_p^2=0.1$,
angles $\xi(0) = \pi/2$ and $\xi(L) = \pi/2 - 0.3\pi$;  
initial polarization 
$(Q/I=0,\ U/I = 1.0,\ V/I = 0.0)$.
}
\label{fig:varyB7}
\end{figure}

\subsection{Transition Probabilities}

Analytic calculations in the adiabatic limit fail for small
frequencies, since in this case the
transition probabilities between instantaneous 
eigenstates are large.
Even in the large frequency regime the adiabatic limit fails unless
the product $gBL >>1$. 
This can be verified explicitly by computing the transition probabilities 
using the procedure discussed in Ref. \cite{ourselves}. The general solution
to the wave equation can be written as
\begin{equation}
|\psi> = \sum_n a_n(z) e^{i\int_0^z dz'\omega_n}|n>
\end{equation}
where $|n>$ and $\omega_n $ are the instantaneous eigenmodes and
eigenfrequencies respectively. The evolution of the coefficients
$a_n(z)$ with $z$ gives an estimate of the transition among different
eigenmodes. These coefficients are obtained by solving the equation
\begin{equation}
\partial_z b_m \approx \sum_{n,n\ne m} b_n
{<m|(\partial_z M)|n>\over \mu_m-\mu_n}
e^{i\int_0^zdz'(\omega_n-\omega_m)}\ ,
\label{eq:bm}
\end{equation}
where we have approximated $\omega_n\approx \omega$ and
$b_m$ is defined by the equation
\begin{equation}
a_m(z)= e^{-{1\over 2}\int_0^z dz' {(\partial_{z'}\omega_m)/
\omega_m}}b_m(z)\approx b_m(z) \label{wkbpre}
\end{equation}

In the small frequency regime the large transition probabilities
are easily understandable: the mass matrix in Eq. \ref{eq:wave_varyB} has two eigenvalues very close to one another. In the opposite limit of large frequencies
we find that the exponent in Eq. \ref{eq:bm} is small. This is because
the exponent is inversely proportional to $\omega$, as long as
$gBL$ is small. The coefficient in
Eq. \ref{eq:wave_varyB} for $m=1$ and $n=2$, for example, in 
this case is found to be proportional to $\xi'$. Integrating this equation
then gives a non-negligible contribution to the transition probability
between different eigenstates. In the limit of large $gBL>>1$ we find 
that the exponent is again large and suppresses the transition probabilities.

\paragraph{Results:}  In Fig. \ref{fig:varyB7} we show a sample of results obtained in the  
case of varying direction of background magnetic field from analytic
calculation in the adiabatic limit as well as
direct numerical integration. Here angle $\xi(0) = \pi/2$ and $\xi(L) = \pi/2 - 0.3\pi$,
i.e. the transverse component of the background magnetic field is
aligned along the y-axis initially and evolves to angle $0.3\pi$
after a distance $L$. The parameters used in this figure are 
 $gB=1.0$, $L=100$, $m_\phi^2/\omega_p^2=0.1$. The initial state of 
 polarization has been chosen such that $Q/I=0,\ U/I=1$, and $V/I=0$. 
The analytic results in this case are in good agreement
with the numerical results, except in the limit of small frequencies.
In the large frequency limit the exponent in Eq. \ref{eq:bm} is 
approximately equal to $igBL/2$. For the parameters chosen this phase
factor is large and hence suppresses the transition probability between
different eigenstates.

\begin{figure}
\psfig{file=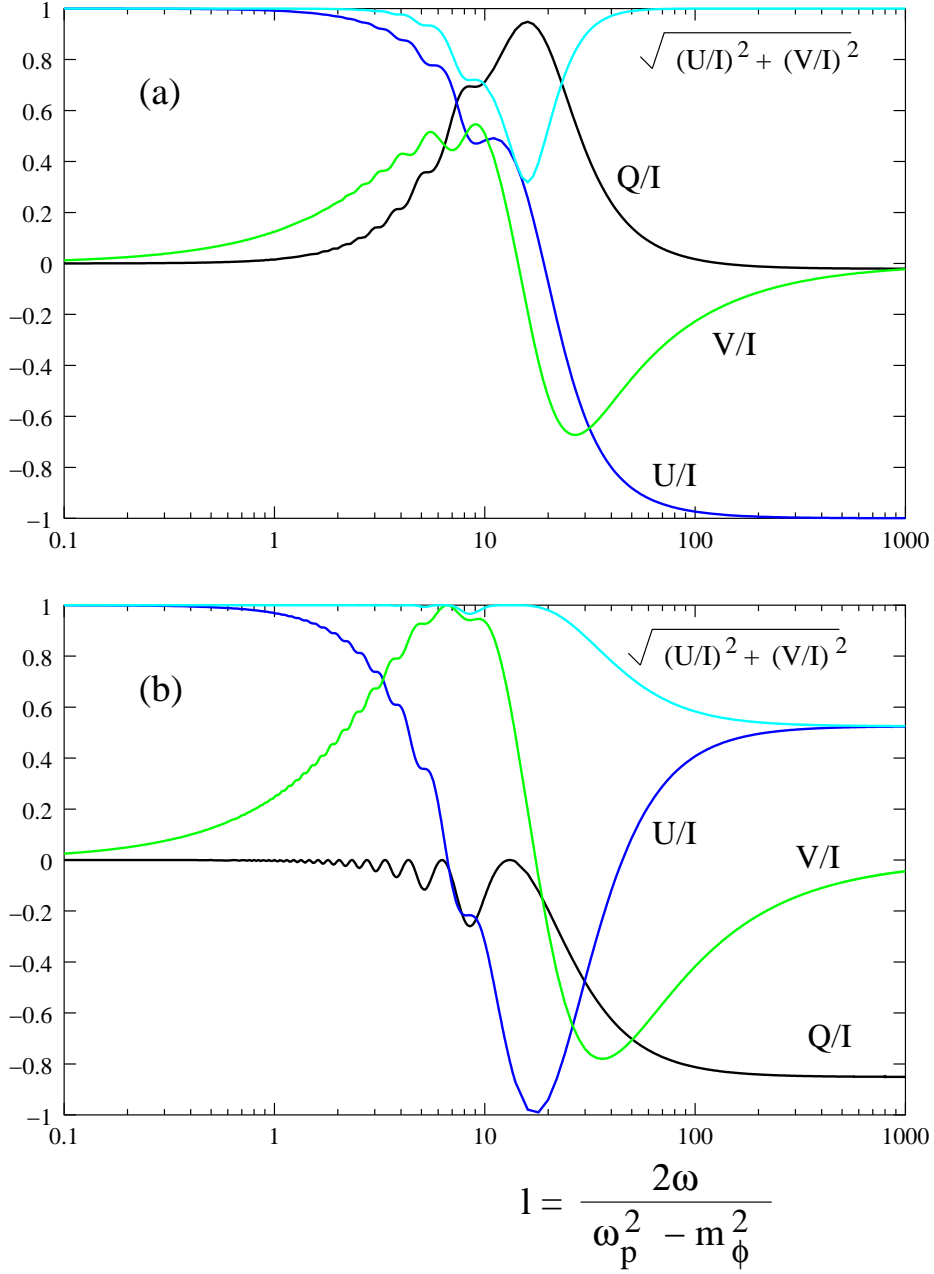}
\caption{(a) Normalized Stokes parameters $Q/I, \,U/I$ and $V/I$ as a function of the length
parameter $l$
for varying background magnetic field direction; the magnitude 
$|\vec {\cal B}|$ and $\omega_{p}$ are constant.   Parameters $gB=0.1$, $L=100$,
$m_\phi^2/\omega_p^2=0.1$; angles $\xi(0) = \pi/2$, $\xi(L) = \pi/2 - 0.3\pi$; 
initial state of the polarization 
$(Q=0,\ U = 1.0,\ V = 0.0)$.  Results for uniform background magnetic field
(b) are shown for comparison.   
}
\label{fig:varyB5}
\end{figure}

\begin{figure}
\psfig{file=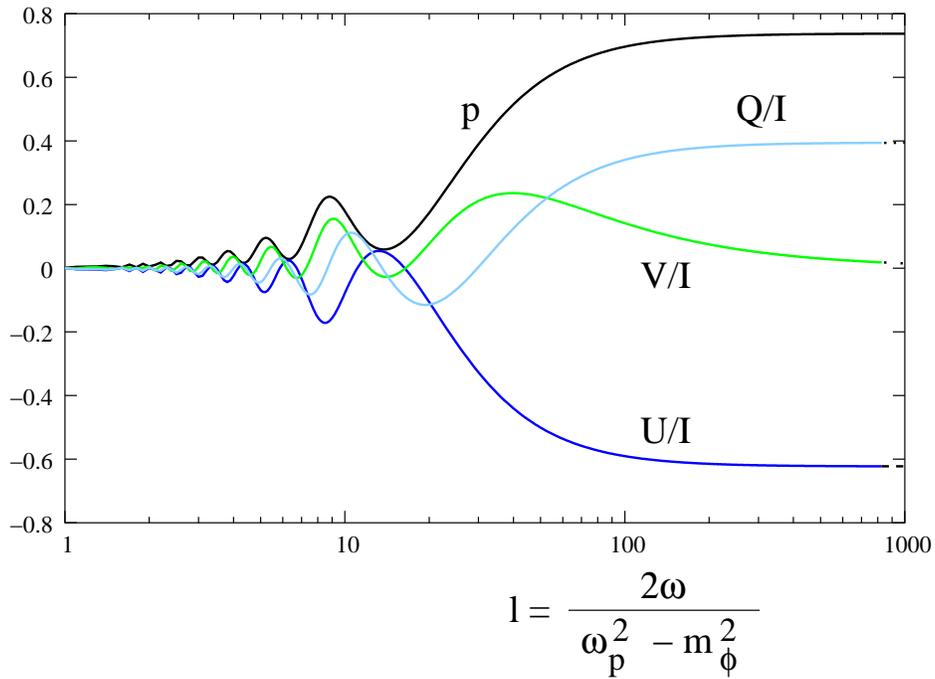}
\caption{The degree of polarization $p$ and the
normalized Stokes parameters $Q/I,U/I$ and $V/I$ as a function of the length
parameter $l$
for varying direction of background magnetic field;
the magnitude $|\vec {\cal B}|$ and $\omega_{p}$ are constant. 
Parameters $gB=0.1$, $L=100$,
$m_\phi^2/\omega_p^2=0.1$; 
 angles $\xi(0) = \pi/2$, $\xi(L) = \pi/2 - 0.3\pi$.  
The wave is assumed to be unpolarized $(Q=0,\ U = V = 0 )$ at source.
}
\label{fig:varyB6}
\end{figure}

In Fig. \ref{fig:varyB5} we show a sample of results obtained in the
case of varying direction of background magnetic field for a smaller 
value of the product $gBL$. Here we choose $gB=0.1$, $L=100$, 
and $m_\phi^2/\omega_p^2=0.1$. In this case we use 
direct numerical integration since the  
analytic results are not reliable. 
The orientation of the background magnetic field is chosen to
be same as in Fig. \ref{fig:varyB7} i.e.
$\xi(0) = \pi/2$ and $\xi(L) = \pi/2 - 0.3\pi$.
  The initial state of 
 polarization has been chosen such that $Q/I=0,\ U/I=1$, and $V/I=0$. 
The results obtained using this parameter choice and with uniform magnetic
field direction are also shown for comparison. 
We find that the
results obtained for the case of varying background magnetic field direction
are considerably different in comparison to what is obtained 
in the case uniform direction. As expected the results agree in the limit
of small $\omega$.  
In Fig. \ref{fig:varyB6} we show the results for the same parameter choice used
in Fig. \ref{fig:varyB5} but with the wave assumed to be unpolarized initially.

\begin{figure}
\psfig{file=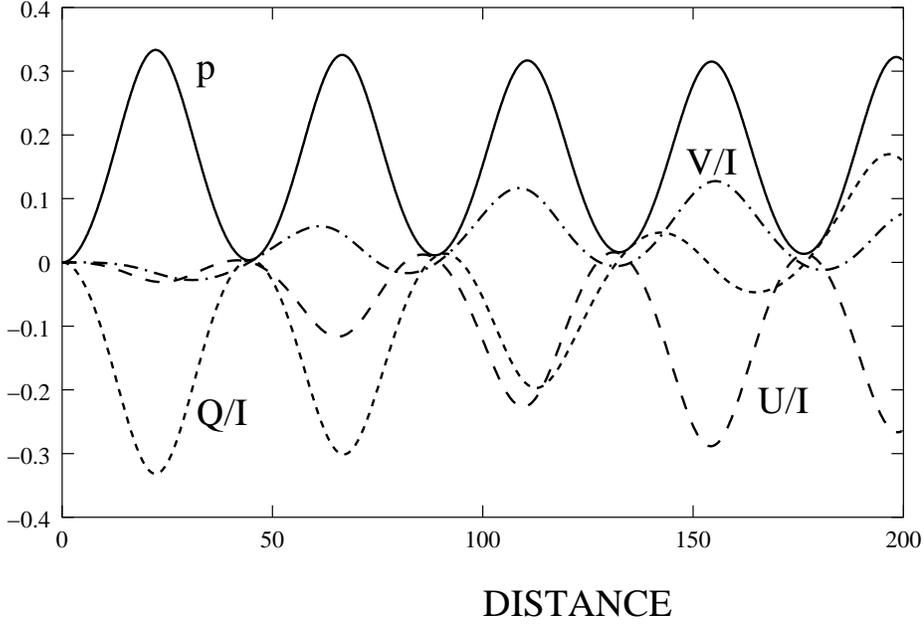}
\caption{The degree of polarization $p$ and the 
normalized Stokes parameters $Q/I,U/I$ and $V/I$ as a function of the distance
of propagation for varying direction of background magnetic field;
the magnitude $|\vec {\cal B}|$ and $\omega_{p}$ are constant. 
The parameters $gB=0.1$, 
$m_\phi^2/\omega_p^2=0.1$,  $l=2\omega/(\omega_p^2-
m_\phi^2) = 10$; 
 angles $\xi(0) = \pi/2$, $\xi(L) = \pi/2 - 0.3\pi$.  
The wave is assumed to be unpolarized $(Q=0,\ U = V = 0 )$ at source.
}
\label{fig:DISTANCE}
\end{figure}

The degree of polarization and the normalized 
Stokes parameters  as a function of distance 
are shown in Fig. \ref{fig:DISTANCE}. Here the parameters are taken to be
same as for the Fig. \ref{fig:varyB5} with the length parameter
$l=2\omega/(\omega_p^2-m_\phi^2)=10$ and the wave is assumed to be unpolarized
at source. We see that all the parameters 
$p,Q/I,U/I,V/I$ oscillate with propagation distance. 

\begin{figure}
\hskip -1.0in
\psfig{file=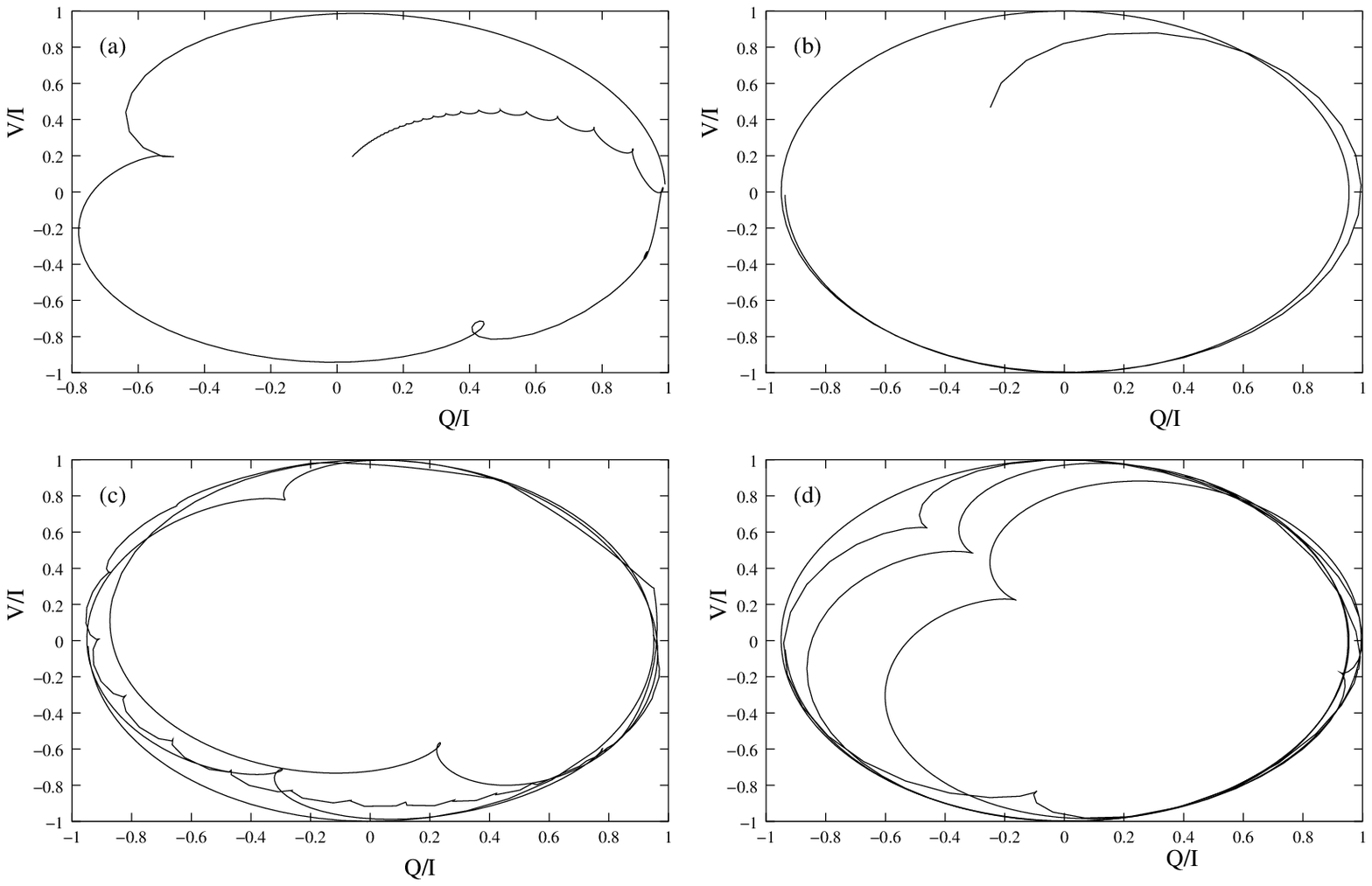}
\caption{A sample of results showing the correlation between the
normalized Stokes
parameters $Q/I$ and $V/I$ for some randomly chosen parameters and initial
state of polarization. The results are shown for  
varying background magnetic field direction with the plasma frequency
and the magnitude of the magnetic field uniform.
Parameters (in arbitrary units) are
(a) $gB = 2,\ L = 10,\  0.04<l<20$, 
(b) $gB = 10,\ L = 10,\  0.4<l<800$, 
(c) $gB = 1,\ L = 50,\  0.2<l<800$ and 
(d) $gB = 10,\ L = 10,\  0.04<l<100$. The ratio $m_\phi^2/\omega_p^2=0.1$;
angles $\xi(0) = \pi/2$ and $\xi(L) = \pi/2 - 0.3\pi$ 
for all the plots.  
 }
\label{fig:cor_B}
\end{figure}

In Fig. \ref{fig:cor_B} we show the relationship between $Q/I$ and $V/I$ for 
several different choice of parameters for the case of varying background
magnetic field. The dependence of $Q/I$ and $V/I$ follows approximately an
elliptical behaviour. This is in contrast to the the case of uniform 
magnetic field direction, which shows such a relationship between
$U/I$ and $V/I$ \cite{ourselves}. 
As in the case of uniform background, 
a simple correlation is seen only for frequencies larger than
a minimum frequency. At low frequencies the relationship becomes 
very complicated.

One may be able to test relationships among different Stokes 
parameters in future observations. To rule out other possible mechanisms affecting data,  certain tests require observations over a sufficiently large frequency interval.

\section{Summary and Conclusion}

The general treatment of mixing of electromagnetic
waves with pseudoscalars in the presence of background magnetic field 
is a surprisingly intricate topic.  The pseudoscalar mixes with (and 
indeed becomes) the longitudinal mode of light, a situation potentially 
generating cumulative deviation compared to treatments assuming the 
fields stay transverse.  Cumulative errors do occur in principle, but 
for parameters of current interest they are fortunately controlled.  
The contribution due to the longitudinal
component can be accommodated by redefining the pseudoscalar
mass parameter $m^2_\phi\rightarrow m^2_\phi + g_\phi^2{\cal B}_L/
\epsilon$.  This simplification led to exploring the problem of
propagation in a magnetic field whose direction may vary along the path.
The condition of adiabaticity is found to be rather stringent:  For
a wide range of parameter space the evolution cannot be assumed to be
adiabatic.

Thus the general problem of mixing of light with pseudoscalars has  
more twists and turns than could have been anticipated early.  Stokes 
parameters show interesting correlations with one another  which are 
distinctively different from those observed for fixed background field 
direction\cite{ourselves}.  Such polarization effects may be
observable with current technology, and may eventually serve either to 
identify new physics, or to put new limits on the  pseudoscalar-photon 
coupling parameters.

\bigskip
\noindent
\bf {Acknowledgments:}
Work supported in part under
Department of Energy grant number DE-FG02-04ER41308.

 \end{document}